\documentclass[11pt]{article}
\usepackage{vietnam,epsfig}

\def\be{\begin{equation}}
\def\ee{\end{equation}}
\def\bea{\begin{eqnarray}}
\def\eea{\end{eqnarray}}

\begin{document}
\vspace*{4cm}
\title{DEPHASING DUE TO A FLUCTUATING\\ FRACTIONAL QUANTUM HALL EDGE CURRENT}

\author{T. K. T. NGUYEN$^{1,2}$, A. CR\'EPIEUX$^{1}$, T. JONCKHEERE$^{1}$, A. V. NGUYEN$^{2}$, Y. LEVINSON$^3$,\\ AND T. MARTIN$^{1}$}

\address{$^1$ Centre de Physique Th\'eorique, CNRS Luminy case 907, 13288 Marseille cedex 9, France \\
$^2$ Institute of Physics and Electronics, 10 Dao Tan, Cong Vi, Ba Dinh, Hanoi, Vietnam\\
$^3$ Department of Condensed Matter Physics, The Weizmann Institute of Science, Rehovot 76100, Israel
}

\maketitle\abstracts{
The dephasing rate of an electron level in a quantum dot, placed next to a fluctuating edge current in the fractional quantum Hall effect, is considered. Using perturbation theory, we first show that this rate has an anomalous dependence on the bias voltage applied to the neighboring quantum point contact, because of the Luttinger liquid physics which describes the fractional Hall fluid.
Next, we describe exactly the weak to strong backscattering crossover using the Bethe-Ansatz solution.}

\section{Introduction}

The presence of electrical environment influences the 
transport through a quantum dot: its energy level
acquires a finite linewidth if the environment has strong charge fluctuations.
Several experiments, performed with a quantum dot embedded in an Aharonov-Bohm loop, probed the phase coherence of 
transport when the dot is coupled to a controlled environment, such as a quantum point contact (QPC) \cite{yacoby1}. Charge fluctuations in the QPC create a fluctuating potential at the dot, modulate 
its electron level, and destroy the coherence of the transmission through the dot \cite{buks_nature}. Theoretical studies for describing this dephasing 
have been developped \cite{levinson_euro39,aleiner}, and were applied to a quantum Hall geometry \cite{levinson_PRB_61}, and  to a normal metal-superconductor QPC \cite{guyon_martin_lesovik}. 
%In all the above, the dephasing rate typically 
%increases when the voltage bias of the QPC is increased.

In this work, we consider the case of dephasing from a QPC in the fractional quantum Hall effect (FQHE)
regime \cite{laughlin}. QPC transmission can be described by tunneling between edge states which represent 
collective excitations of the quantum Hall fluid. 
It is interesting because the transport properties deviate strongly from the case of normal conductors \cite{kane_PRL92,kane_PRL94,chamon_PRB95}: 
for the weak backscattering (BS) case, the current at zero temperature may increase when the voltage bias is lowered, 
while in the strong BS case, the $I(V)$ is highly non linear. 
It is thus important to address the issue of dephasing from a Luttinger liquid. 
%Here, we consider the case of simple Laughlin fractions, with filling factor $\nu=1/m$ ($m$ odd integer).  
%The dephasing of a state in the dot is induced by its capacitive coupling to the biased QPC, assuming that the level modulation in the dot is a Gaussian process, and neglecting back-action effects.

\section{Dephasing in the fractional quantum Hall regime}

\begin{figure}[h]
\centerline{\includegraphics[width=6cm]{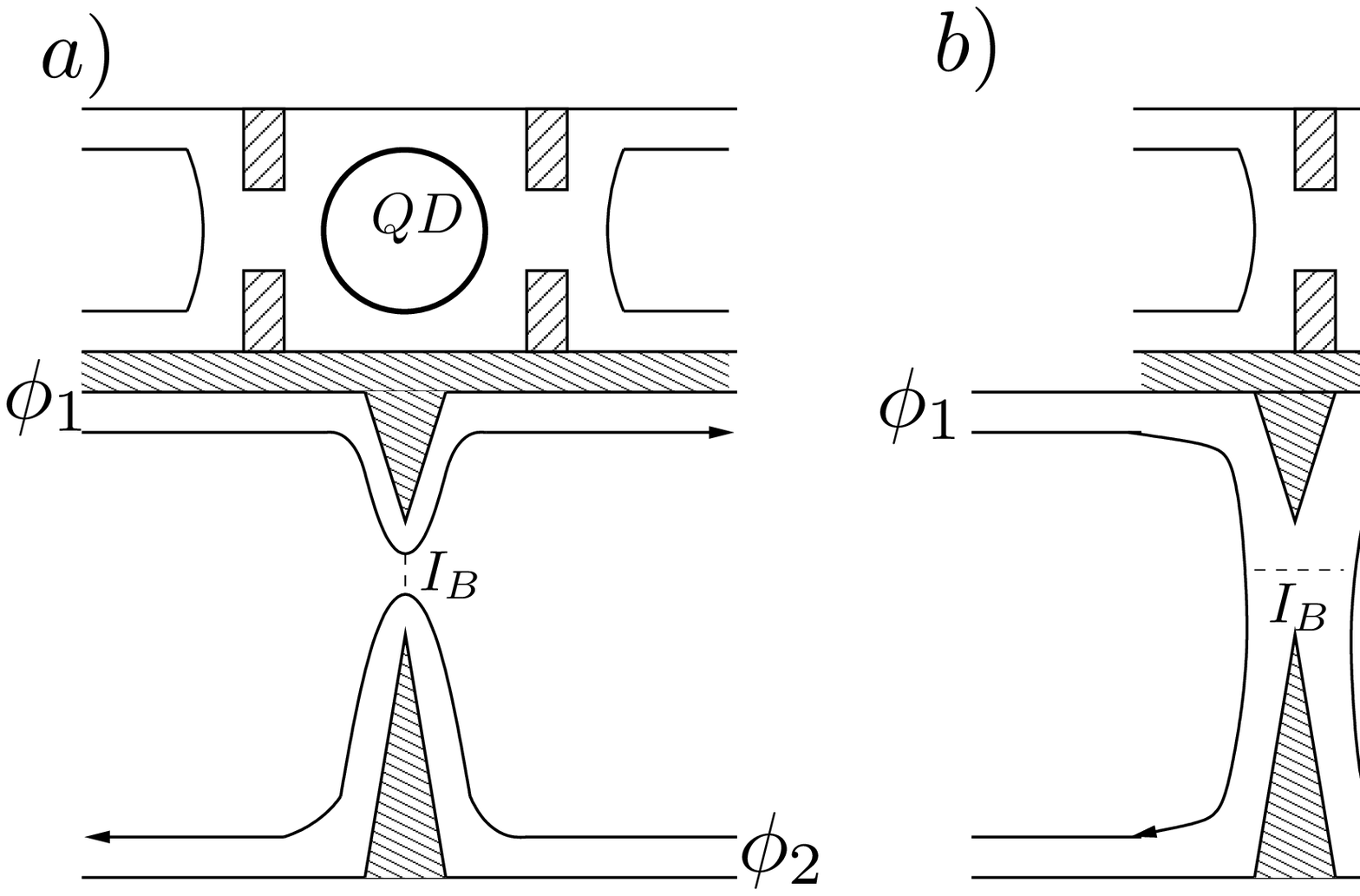}}
\caption{Schematic description of the setup: the quantum dot (top) is coupled capacitively to a quantum point contact in 
the FQHE regime: a) case of weak backscattering, b) case of strong backscattering.}
\label{fig1}
\end{figure}

The system we consider is depicted in Fig.~\ref{fig1}. 
The single level Hamiltonian for the dot
reads $H_{QD}=\epsilon_0 c^\dag c$, where $c^\dag$ creates an electron.
This dot is coupled capacitively to a point contact in the FQHE. 
The Hamiltonian which describes the edge modes in the absence of tunneling is: 
$H_{0}=(\hbar v_F/4\pi)\int
dx[(\partial_x\phi_1)^2+(\partial_x\phi_2)^2]$,
where $\phi_1$ and $\phi_2$ are the chiral Luttinger bosonic fields, which are related to the electron density operators $\rho_{1(2)}$ by $\partial_x\phi_{1(2)}(x)=\pi\rho_{1(2)}(x)/\sqrt{\nu}$.\\

%%%%%%%%%%%%%%%%%%%%%%%%%%%%%%%%%%%%%%%%%%%%%%%%%%%%%%%
%%%%%%%%%%%%%%%%%%%%%%%%%%

By varying the gate potential of the QPC, one can switch from a weak BS situation, 
where the Hall liquid remains in one piece (Fig.~\ref{fig1}a), to a strong BS situation where the Hall liquid is split in two
(Fig.~\ref{fig1}b). In the former case, the entities which tunnel are edge quasiparticle excitations. In the latter case, 
between the two fluids, only electrons can tunnel. We consider the weak BS case, and then we use a 
duality transformation \cite{kane_PRL92,chamon_PRB96} to describe the strong BS case. 
The tunneling Hamiltonian between edges 1 and 2 reads
$H_t=\Gamma_0 e^{i\omega_0t}\psi^{+}_{2}(0)\psi_{1}(0)+h.c.$
where we have used a Peierls substitution to include the voltage: $\omega_0=e^\star V/\hbar$ ($e^\star=\nu e$ 
is the effective charge, $\nu$ is the filling factor).
The quasiparticle operator is $\psi_i(x)=e^{i\sqrt{\nu}\phi_i(x)}/\sqrt{2\pi\alpha}$
(the spatial cutoff is $\alpha=v_F\tau_0$, with $\tau_0$ the temporal cutoff).

The Hamiltonian describing the interaction between the dot and the QPC reads $H_{int}=c^{+}c\int dx f(x)\rho_1(x)$, with $f(x)$ a Coulomb interaction kernel, which is assumed to include screening by the nearby gates $f(x)\simeq e^2 e^{-|x|/\lambda_s}/\sqrt{x^2+d^2}$, where $d$ is the distance from the dot to the edge, $\lambda_s$ is a screening length. 
%The dephasing of an electron state in a dot coupled to a fluctuating current is caused by the electron density fluctuations, 
%which generate a fluctuating potential in the dot, resulting in a blurring of the energy level $\epsilon_0$. 
The dephasing rate, expressed in terms of irreducible charge fluctuations in the adjacent wire, is written as \cite{levinson_euro39,aleiner,levinson_PRB_61}: 
\begin{eqnarray}
\tau_{\varphi}^{-1}&=&\frac{1}{4\hbar^2}\int^{\infty}_{-\infty}dt\int\ dx f(x)\int
dx^\prime f(x^\prime)\langle\langle\rho_1(x,t)\rho_1(x^\prime,0)+\rho_1(x^\prime,0)\rho_1(x,t)\rangle\rangle~.
\end{eqnarray}
%In normal and superconducting systems, the dephasing rate can be calculated using the scattering approach.
%For Luttinger liquids and in particular for the FQHE, it is conveninent to use the Keldysh approach \cite{chamon_PRB95,martin_noise}.

%A tunneling event (at $x=0$) creates an excitation which need to propagate to the location of the dot.
The equilibrium contribution to the dephasing rate corresponds to the zero order in the tunneling amplitude $\Gamma_0$:
\begin{eqnarray}
(\tau_{\varphi}^{-1})^{(0)}&=&\frac{\nu}{4\pi^2\hbar^2}\int^{\infty}_{-\infty}dt \int dx f(x)\int dx^\prime f(x^\prime)\sum_{\eta=\pm}\partial_{xx^\prime}^{2}G_{1}^{\eta-\eta}(x-x^\prime,t)~, 
\label{A0}
\end{eqnarray}
where the bosonic Green's function is $ G_{i}^{\eta_1\eta_2}(x-x',t_1-t_2)=\langle \phi_{i}(x,t_{1}^{\eta_1})\phi_{i}(x',t_{2}^{\eta_2})-\phi_{i}^2\rangle$. 
The coefficients $\eta$,$\eta_{1,2}=\pm$ identify the upper/lower branch of the Keldysh contour. There is no contribution to first order in the tunneling Hamiltonian, while the non-equilibrium contribution corresponding to the second order in $\Gamma_0$ exists:
\begin{eqnarray}
&&(\tau_{\varphi}^{-1})^{(2)}=-\frac{\nu}{4\pi^2\hbar^4}\frac{\Gamma_0^2}{2(2\pi\alpha)^2}\int^{\infty}_{-\infty}dt \int dx f(x)\int dx^\prime f(x^\prime) \sum_{\eta,\eta_1,\eta_2,\epsilon}\eta_1\eta_2\int^{\infty}_{-\infty}dt_1\int^{\infty}_{-\infty}dt_2 \nonumber\\
&&\times e^{i\epsilon\omega_0(t_1-t_2)}e^{\nu G_{2}^{\eta_1\eta_2}(0,t_1-t_2)}e^{\nu G_{1}^{\eta_1\eta_2}(0,t_1-t_2)} \left\{\partial_{xx^\prime}^{2}G_{1}^{\eta-\eta}(x-x^\prime,t)\right.\nonumber\\
&&+\left.\nu[\partial_x G_{1}^{\eta\eta_1}(x,t-t_1)-\partial_x G_{1}^{\eta\eta_2}(x,t-t_2)] [\partial_{x^\prime}G_{1}^{-\eta\eta_1}(x^\prime,-t_1)-\partial_{x^\prime}G_{1}^{-\eta\eta_2}(x^\prime,-t_2)]\right\}~.
\label{A2}
\end{eqnarray}

The dephasing rate depends on the geometry of the set up via the length scales d, $\lambda_s$, and $\alpha$. The assumption of strong screening $\lambda_s\sim\alpha=v_F\tau_0$ is made ($f(x)\simeq 2 e^2\alpha\delta(x)/d$). 
Inserting
$G^{\eta\eta^\prime}_{1}(x,t)=-\ln\Big\{\sinh[\pi[(x/v_F-t)((\eta+\eta^\prime){\rm  sgn}(t)-(\eta-\eta^\prime))/2+i\tau_0]/\hbar\beta]\Big/\sinh[i\pi\tau_0/\hbar\beta]\Big\}$,
 where $\beta=1/k_B T$,
in the dephasing rate gives:
$(\tau_{\varphi}^{-1})^{(0)}=4e^4\tau_0^2\nu/\pi\hbar^3\beta d^2$ and,
\begin{equation}
(\tau_{\varphi}^{-1})^{\!(2)}=\frac{e^4\Gamma_0^2}{\pi^2\hbar^4 v_F^2 d^2}\frac{\nu^2\tau_0^{2\nu}}{\Gamma(2\nu)}\left(\frac{2\pi}{\hbar\beta}\right)^{2\nu-1}\cosh\left(\frac{\omega_0\hbar\beta}{2}\right)\left|\Gamma\left(\nu+i\frac{\omega_0\hbar\beta}{2\pi}\right)\right|^{2}~.
\label{gamma_nu}
\end{equation}

%In the zero temperature limit, we have $(\tau_{\varphi}^{-1})^{(0)}=0$ and, 
%\begin{equation}
%(\tau_{\varphi}^{-1})^{(2)}=\frac{e^4\Gamma_0^2}{\pi\hbar^4 v_F^2 d^2}\frac{\nu^2\tau_{0}^{2\nu}}{\Gamma(2\nu)}|\omega_0|^{2\nu-1}~.
%\label{a2T0}
%\end{equation}

Note that $(\tau_{\varphi}^{-1})^{(2)}=(e\tau_0/d)^2 S_I(0)$, with $S_I(0)=\int dt \langle\langle I(t)I(0)\rangle\rangle$ the zero-frequency BS noise. The non-equilibrium contribution of the dephasing rate is proportional to the zero-frequency noise \cite{kane_PRL94,chamon_PRB95,chamon_PRB96,martin_noise} in the quantum Hall liquid. At zero temperature, the non-equilibrium dephasing rate given by Eq.~(\ref{gamma_nu}) leads to
$(\tau_{\varphi}^{-1})^{(2)}\propto|\omega_0|^{2\nu-1}$
and depends on the QPC bias with the exponent $2\nu-1<0$, in sharp contrast with the linear dependence obtained by Levinson \cite{levinson_euro39}.

\vspace*{0.5cm}

%%%%%%%%%%%%%%%%%%%%%%%%%%%%%%%%%%%%%%%%%%%%%%%%%%%%%%%
\begin{figure}[h]
\centerline{\includegraphics[width=5cm, angle=-90]{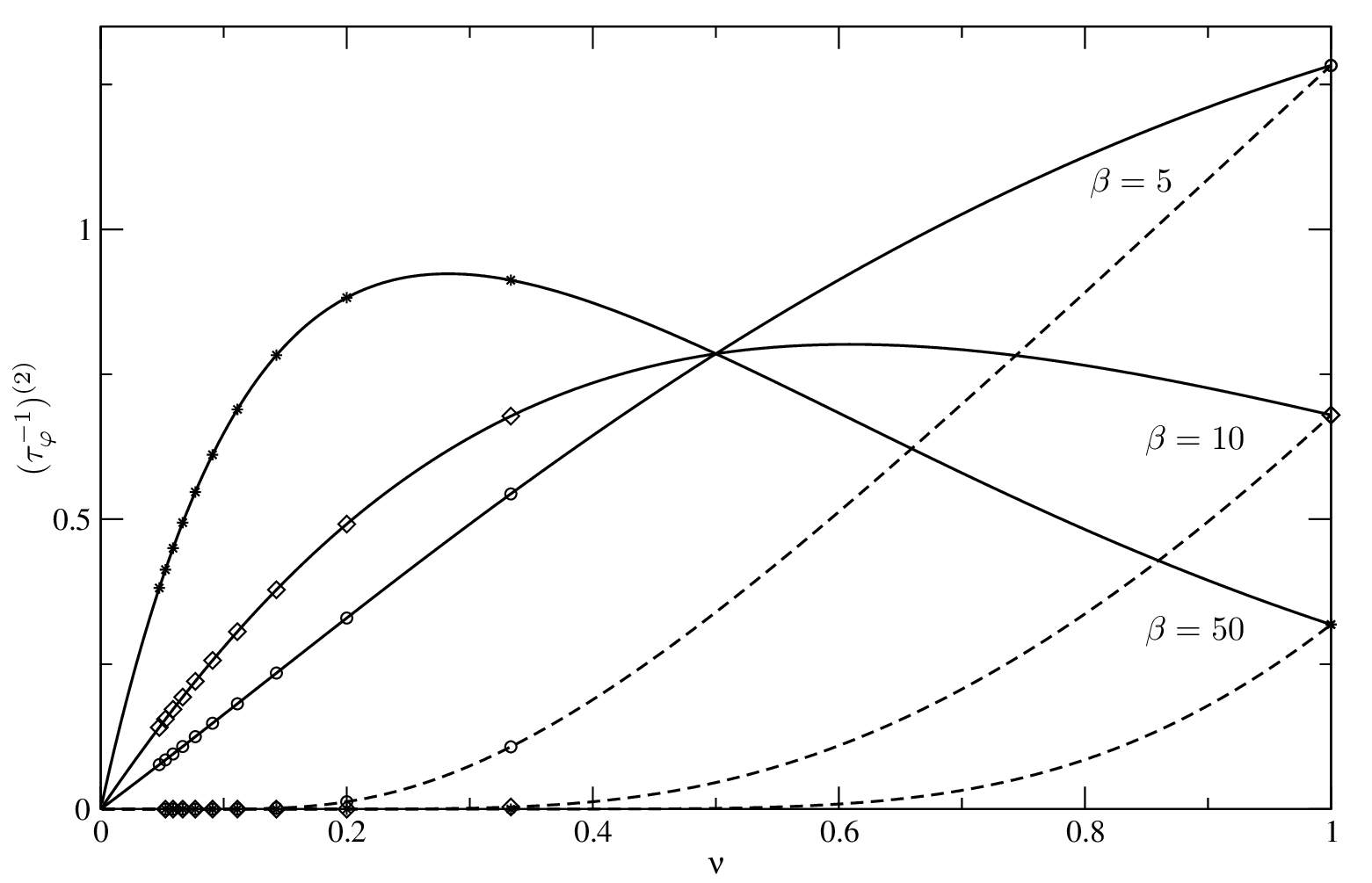}
\hspace{0.2cm}
\includegraphics[width=5cm, angle=-90]{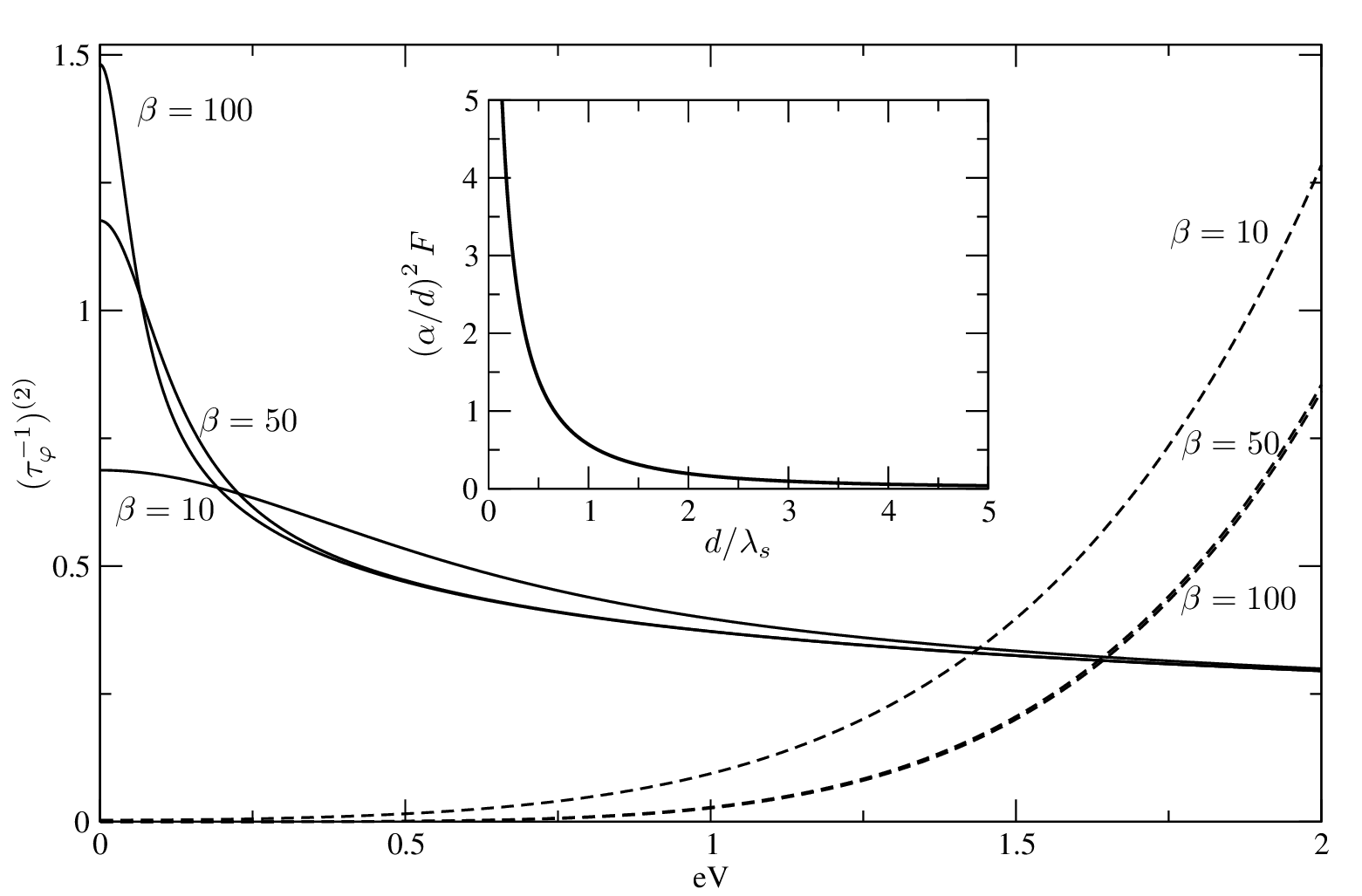}}
\caption{(Left) Dephasing rate, plotted in units of $e^4\Gamma^2_0\tau_0/\pi^2\hbar^4 v_F^2 d^2$, as a function of the filling factor for weak BS (full line) and strong BS (dashed line) at QPC bias $eV=0.1$. 
The star, diamond and circle points correspond to the Laughlin fractions $\nu=1/m$, $m$ odd integer. (Right) Dephasing rate as a function of QPC bias with $\nu=1/3$ for weak BS (full line) and strong BS (dashed line). The insert is the ratio of non-equilibrium contribution in dephasing rate between the arbitrary screening and strong screening multiplied by $(\alpha/d)^2$ as a function of $d/\lambda_s$.}
\label{fig2}
\end{figure}
%%%%%%%%%%%%%%%%%%%%%%%%%%

In the left of Fig.~\ref{fig2}, we plot the dependence of the non-equilibrium contribution of the dephasing rate on the filling factor $\nu$ for weak BS and strong BS for several 
temperatures ($\beta=5, 10, 50$) at fixed QPC bias. $\nu$ is considered as a continuous variable, while it has physical meaning only at Laughlin fractions \cite{laughlin}. For the strong BS case, the dephasing rate increases when the $\nu$ increases. For weak BS and $1/\beta\ll eV$, the dephasing rate has a local maximum at $\nu<1/2$, the position of which depends on temperature: when the temperature increases, it gets closer to $\nu=1/2$.
The rate at $\nu=1$ is smaller than that at $\nu=1/3$. This result demonstrates that for two different filling factors, we can have comparable dephasing rates. For weak BS and $1/\beta > eV$, the dephasing rate increases when the filling factor increases. In the right of Fig.~\ref{fig2}, the dependence of the dephasing rate on the QPC bias voltage is plotted for $\nu=1/3$ and several temperatures. In the case of strong BS, the dephasing rate increases when the bias increases.  For $1/\beta\ll eV$, the dephasing rate saturates, whereas for $1/\beta > eV$, the dephasing rate increases when $eV$ increases, but it increases from a finite value (not shown), which is proportional to the temperature. Things are quite different at weak BS. At high temperatures, the dephasing rate decreases  
when we increase $eV$: this behavior is symptomatic of current and noise characteristic in a Luttinger liquid. In the low temperature case $1/\beta\ll eV$, for small $eV$, the lower the temperature, the bigger the dephasing rate and the faster it decreases when we increase $eV$. 

%At $T=0$, the dephasing rate is ``infinite'' at $eV=0$. This Luttinger liquid behavior is in sharp contrast with the result of Levinson \cite{levinson_euro39}. 

%We find that the dephasing rates evaluated at different temperatures coincide at the (unphysical) value $\nu=1/2$, because the hyperbolic cosine multiplied by the squared modulus of the Gamma function in Eq.~(\ref{gamma_nu}) does not depend on temperature, 
%while at the same time the exponent $(2\nu -1)$ is zero: this is known for perturbative calculations of the backscattering current and noise.

\section{General formula for the decoherence rate}

The charge fluctuations are directly related to the current fluctuations along the edges which are identical to the fluctuations of the tunneling current. The tunneling current fluctuations were computed non pertubatively using Bethe-Ansatz techniques \cite{fendley_saleur}. We can therefore invoke current conservation at the point contact to derive a general formula for the decoherence rate, which describes the crossover from weak to strong BS: 
$(\tau_{\varphi}^{-1})^{(2)}=(e^3\tau_0^2/d^2)(V G_{diff}-I)\nu/(1-\nu)$
where $G_{diff}=\partial_V I$ is the differential conductance and $I$ is the current \cite{fendley}. This expression allows us to describe 
the crossover in the dephasing rate from the weak to the strong BS regime. 

Remarkably, it is possible to go beyond the strong screening limit, and one can 
compute Eq.~(\ref{A2}) for an arbitrary Coulomb kernel $f(x)$. The result can be displayed in terms of the 
ratio between the arbitrary screening dephasing rate and the strong screening dephasing rate: 
\begin{equation}
F\equiv\frac{(\tau_{\varphi}^{-1})^{(2)}}{(\tau_{\varphi}^{-1})^{(2)}_{\lambda_s\rightarrow \alpha}}=\frac{d^2}{(e\alpha)^2}\left[\int_{0}^{\infty}\!\!dx f(x)\right]^2~.
\label{arbitrary_screening}\end{equation}

%where the integral is a function of $d/\lambda_s$, and we recall that $\alpha$ is the spatial cutoff. 
If the Coulomb interaction kernel $f(x)$ is chosen as
suggested before, the dephasing rate at arbitrary $\lambda_s$ 
has an analytical expression:  
$F=(\pi d/2\alpha)^2 [E_0(d/\lambda_s)+N_0(d/\lambda_s)]$, where $E_0(d/\lambda_s)$ and  $N_0(d/\lambda_s)$ are the Weber and 
the Neumann functions of zero order. $F$ is plotted in the insert of Fig.~\ref{fig2}:
$F$ is infinite in the absence of screening. However, the presence of metallic gates
always imposes a finite screening length. $F$ decreases with $d/\lambda_s$ and approaches $1$ when $\lambda_s$ is 
close to the spatial cutoff $\alpha$. The dephasing rate 
increases when the screening decreases.      

\section{Conclusion}

We have established a general formula for the dephasing rate of a quantum dot located 
in the proximity of a fluctuating fractional edge current. 
For strong screening, we have shown that the dephasing rate is given by the tunneling current noise, 
for both weak and strong BS. For weaker screening, the spatial 
dependence of the density-density correlation function has to be taken into account,
but we have shown explicitly that the long range nature of the Coulomb interaction can be 
included as a trivial multiplicative factor. The fact that the dephasing rate decreases with increasing 
voltage can be reconciled with the fact that the charge noise is directly related to the BS 
current noise in the FQHE. There it is known, and seen experimentally, that 
when the  bias voltage dominates over the temperature, both 
the tunneling current and noise bear a power law 
dependence $\sim V^{2\nu-1}$ with a negative exponent. The fact that at low temperatures, 
the dephasing rate for filling factors can be lower than that of the integer quantum Hall effect 
comes as a surprise and is a consequence of chiral Luttinger liquid theory.

%The present results could be tested with gated heterostructures as in Ref. \cite{sprinzak},
%provided that the electron mobility and the magnetic field are further increased in order 
%to achieve the FQHE regime and provided that the quantum dot is placed next to the QPC as in Fig.~1. 


\section*{References}
\begin{thebibliography}{99}
\bibitem{yacoby1} A. Yacoby {\it et al.}, Phys. Rev. Lett. {\bf 74}, 4047 (1995); A. Yacoby {\it et al.}, Phys. Rev. B {\bf 53}, 9583 (1996); E. Buks {\it et al.}, Phys. Rev. Lett.  {\bf 77}, 4664 (1996); R. Schuster {\it et al.}, Nature {\bf 385}, 417 (1997). 

\bibitem{buks_nature} E. Buks {\it et al.}, Nature {\bf 391}, 871 (1998); D. Sprinzak {\it et al.}, Phys. Rev. Lett. {\bf 84}, 5820 (2000).

\bibitem{levinson_euro39} Y. Levinson, Europhys. Lett. {\bf 39}, 299 (1997).
%6
\bibitem{aleiner} I.L. Aleiner, N.S. Wingreen, and Y. Meir, Phy. Rev. Lett. {\bf 79}, 3740 (1997).
%7
\bibitem{levinson_PRB_61} Y. Levinson, Phys. Rev. B {\bf 61}, 4748 (2000).
%8
\bibitem{guyon_martin_lesovik} R. Guyon, T. Martin, and G.B. Lesovik, Phys. Rev. B {\bf 64}, 035315 (2001).

\bibitem{laughlin} 
D.C. Tsui {\it et al.},
Phys. Rev. Lett. {\bf 48}, 1559 (1982);
R.B. Laughlin, {\it ibid.} {\bf 50}, 1395 (1983).

%\bibitem{wen_92} X. G. Wen, Int. J. Mod. Phys. B {\bf 6}, 1711 (1992).

\bibitem{kane_PRL92} C.L. Kane and M.P.A. Fisher, Phys. Rev. Lett. {\bf 68}, 1220 (1992).

\bibitem{kane_PRL94} C.L. Kane and M.P.A. Fisher, Phys. Rev. Lett. {\bf 72}, 724 (1994).

\bibitem{chamon_PRB95} C. Chamon, D.E. Freed, and X.G. Wen, Phys. Rev. B {\bf 51}, 2363 (1995).

\bibitem{chamon_PRB96} C. Chamon, D.E. Freed, and X.G. Wen, Phys. Rev. B {\bf 53}, 4033 (1996).

\bibitem{martin_noise} T. Martin in Les Houches Summer School session LXXXI, 
edited by E. Akkermans, H.~Bouchiat, S. Gu\'eron, and G. Montambaux (Elsevier, 2005).

%\bibitem{guyon} R. Guyon and T. Martin, unpublished.

%\bibitem{sami} L. Saminadayar, D.C. Glattli, Y. Jin, and B. Etienne, Phys. Rev. Lett. {\bf 79}, 2526 (1997).

%\bibitem{picciotto} R. de-Picciotto {\it et al.}, Nature {\bf 389}, 162 (1997);
%M. Reznikov {\it et al.}, Nature {\bf 399}, 238 (1999).

\bibitem{fendley_saleur} P. Fendley and H. Saleur, Phys. Rev. B {\bf 54}, 10845 (1996).

\bibitem{fendley} P. Fendley {\it et al.}, Phys. Rev. Lett. {\bf 75}, 2196 (1995); Phys. Rev. B {\bf 52}, 8934 (1995).
\end{thebibliography}
\end{document}